\documentclass[preprint,showpacs,preprintnumbers,amsmath,amssymb]{revtex4-1}
\usepackage{dcolumn}
\usepackage{bm}
\usepackage{epsfig,amssymb}
\newcommand{\sech}{\mbox{\rm sech}}

\begin{document}
\title{Solitary waves in three-dimensional crystal-like structures}
\author{Edward Ar\'evalo}
\affiliation{Technische Universit\"at Darmstadt, Institut f\"ur
  Theorie elektromagnetischer Felder, TEMF, Schlo{\ss}gartenstr. 8
  D-64289 Darmstadt, Germany}
\date{\today}

\begin{abstract}
The motion of three-dimensional (3D) solitary waves and vortices in
nonlinear crystal-like structures, such as photonic materials, is
studied. It is demonstrated that collective excitations in these systems can
be tailored to move in particular directions of the 3D system. The
effect of modulation instability is studied
showing that in some cases it can be delayed by using a lensing
factor. Analytical results supported by numerical 
simulations are presented. 
\end{abstract}

\maketitle

\section{Introduction}

Recently, great progress has been made in the
experimental fabrication of three-dimensional (3D) crystal-like
optical nano structures, such as photonic crystals
\cite{siliconCrystal09} and  photonic metamaterials
\cite{natMatLett08}. Light in 3D crystal-like optical 
nano structures can interact with the regular pattern of the structure
setting up 
resonances. These resonances can cause beams and pulses to be
deflected in unconventional directions and even to slow down the speed
of light. Engineered optical metamaterials
with unique electromagnetic properties, have become in recent years a
hot research topic because 
of their interesting physics and exciting potential applications as,
e.g cloaking \cite{cloakingE2008,cloakingM2008}, or negative
refractive index \cite{metamat06}, among others.

The theoretical analysis of these  
systems have been mostly performed with the help of numerical
simulations, where the 3D spatial distribution of the effective
electromagnetic properties of the material medium (such as the
permittivity and permeability) are tailored to have specific
electromagnetic properties in the continuum limit
\cite{cloakingE2008}. Other discrete 
effects of the system on the propagation of light waves are usually
neglected. Similar approximations have been also adopted for
studying  cloaking of matter waves \cite{cloakingM2008}. So far, most
of the studies have been done for linear planewaves, so nonlinear
effects have been neglected. There is, however,  strong evidence from
low-dimensional 
systems \cite{microresonators03,cavitySoliton,quadartic2007,edward09-2D}
that discrete effects in combination with  nonlinearities 
may play an important role in describing  realistic 3D crystal-like structures. 
So,  nonlinear excitations such as moving solitary waves (in short
solitons) and vortices can be expected.

With respect to the theoretical models, it has 
been shown that electromagnetic waves interacting with metamaterials
in the tight binding limit can be described as a 3D lattice of
microresonators modeled by the 3D discrete nonlinear Schr\"odinger
equation (3D-DNLSE) \cite{metamat06}. In the case of 3D photonic crystals,
so far, no discrete model has been proposed. However, 
it is well known that the  dynamics of light beams in
photonic waveguides can be described  by the 2D-DNLSE
\cite{quadartic2007,edward09-2D}, where the
discreteness is transversal to the light propagation. Moreover,
it has been also shown that trapped light waves travelling
along chains of optical resonators can be described by the 1D DNLSE
\cite{microresonators03,cavitySoliton}, where the discreteness is
longitudinal to the light propagation. These two
low-dimensional models  strongly
suggest that 3D photonic crystals with high index of refraction
\cite{natMatLett08,siliconCrystal09} may be  effectively modeled in
the tight binding limit by a 3D lattice of optical resonators governed
by the  3D-DNLSE \cite{diammods04,carretero05,3ddnls09}. 

Notice that the 3D-DNLSE 
is an ubiquitous dynamical-lattice
model which may emerge from a variety of other important problems and has its
direct physical realization in Bose-Einstein 
condensates (BECs) trapped in strong optical lattices
\cite{diammods04,carretero05,3ddnls09}.

The aim of the present work is to study the dynamics of moving
solitons and vortices in nonlinear 3D-crystal-like structures
described by the  3D-DNLSE.

\begin{figure}
\centerline{\epsfxsize=7truecm \epsffile{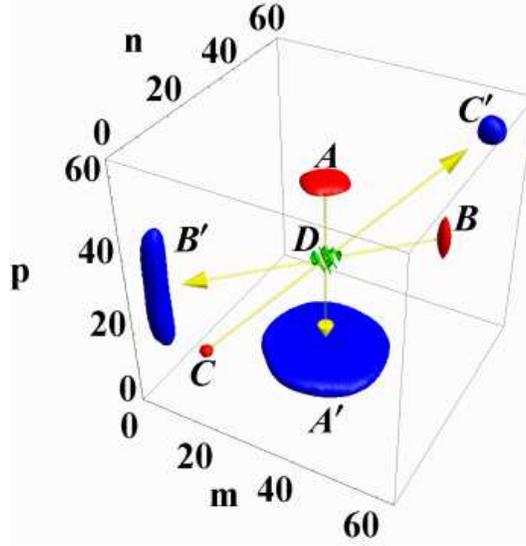}}
\caption{(Color online) Superposition of snapshots of
  $\rho_{m,n,p}$ isosurfaces at different time values of a
  soliton-soliton-soliton collision.  
The isosurfaces are defined as $\kappa \rho_{max}(t)$, where
the maximum $\rho_{max}(t)
=\operatorname*{max}_{m,n,p}\, \rho_{m,n,p}(t)$, and $\kappa=0.5$.
The arrows (in yellow) are meant to guide the eye and show the
path and direction of motion of the solitons. At $t=0$
[labels A, B, and C (red color)] the initial conditions
follow from  Eq. (\ref{solsolution1}), where (A) ${\bf k}=0.95
\frac{\pi}{2}\{0,0,-1\}$, ${\bf r}_{m,n,p}=\{32,32,52\}$, $\Omega_0=1.05$, (B) ${\bf k}=0.95
\frac{\pi}{2}\{-1,-1,0\}$, ${\bf r}_{m,n,p}=\{52,52,0\}$, $\Omega_0=1.1$, and (C) ${\bf k}=0.95
\frac{\pi}{2}\{1,1,1\}$, ${\bf r}_{m,n,p}=\{12,12,12\}$, $\Omega_0=1.5$. At $t=10$ [label D (green
  color)] the collision can be observed. And at $t=24$ [labels $\rm
  A^{\prime}$, $\rm B^{\prime}$, and $\rm C^{\prime}$ (blue color)]
the solitons after the collision can be observed. Other parameters are $U=-1$,
$J_x=J_y=J_z=1$, $\gamma_x=\gamma_y=\gamma_z= 1/3$. Here,
$\beta_x=\beta_y=\beta_z=+1$, so only sech-type solutions
in Eq. (\ref{solsolution1}) are
considered.}  
\label{fig1}
\end{figure}
\begin{figure}
\centerline{\epsfxsize=6truecm \epsffile{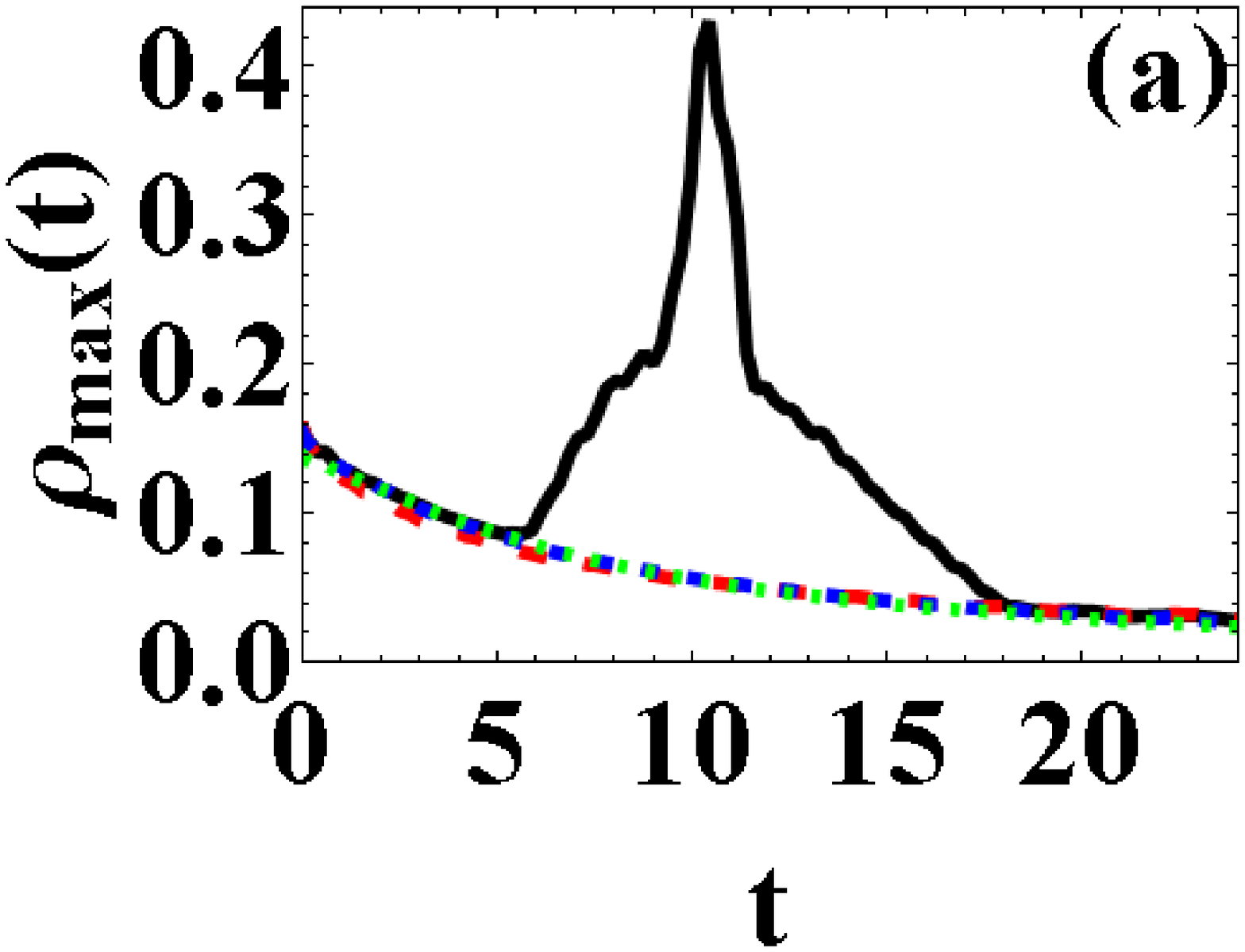}}
\centerline{\epsfxsize=6truecm \epsffile{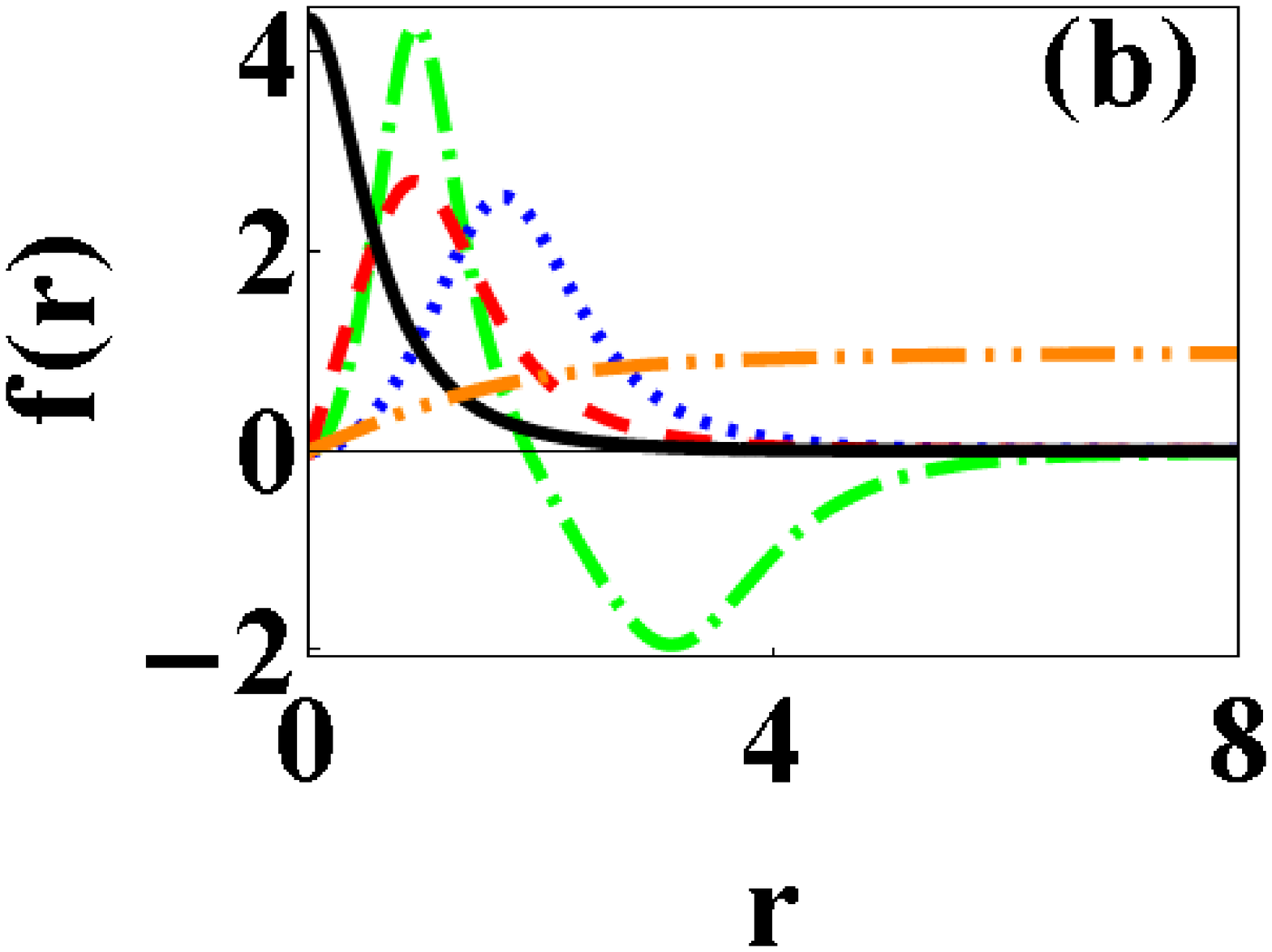}}
\caption{(Color online) a: Evolution of the maximum $\rho_{max}(t)
=\operatorname*{max}_{m,n,p}\, \rho_{m,n,p}(t)$ 
for the collision in Fig. \ref{fig1} (Black solid line). For
comparison, $\rho_{max}(t)$ of the individual solitons (cases where
the solitons do not collide) plotted in Fig. \ref{fig1} is presented
(red, blue, and green dashed lines; lines lie very near each other). b:
some solutions of the Eq. (\ref{spherical2}) for the cases $\beta=1$
[$\nu=0$, (black solid line), $\nu=1$ (red dashed line), $\nu=2$
(blue dotted line and green dot-dash line)], and  $\beta=-1$ [$\nu=1$
  (orange double-dot-dash line)].}  
\label{fig2}
\end{figure}

\section{Theory}

The general form of the 
3D-DNLSE with coupling constants $J_x$, $J_y$ and $J_z$ is
\begin{eqnarray}\label{dnls1}
&&i\partial_t \psi_{m,n,p} +J_x(\psi_{m-1,n,p}+\psi_{m+1,n,p})\nonumber\\
&&+ J_y(\psi_{m,n-1,p}+\psi_{m,n+1,p})+ J_z(\psi_{m,n,p-1}+\psi_{m,n,p+1})\nonumber\\
&&-U\,\rho_{m,n,p}\,\psi_{m,n,p}=0,
\end{eqnarray}
where 
\begin{equation}
\rho_{m,n,p}=\vert \psi_{m,n,p}\vert^2
\end{equation}
 can be interpreted as a
intensity (e.g. in crystals built of microresonators)
or as a probability-density (in BEC arrays).
In Eq. (\ref{dnls1}) the nonlinear coefficient $U$ is a real constant and $t$
is the time coordinate.

In order to proceed we shall consider an expansion into
harmonics of a travelling wave ansatz for an envelope complex
function, i.e
\begin{equation}\label{ansatz1}
\psi_{m,n,p}=\sum\limits_{\mu,\nu,\xi=1}^\infty
\chi_{\mu,\nu,\xi}({\bf S}_{\mu,\nu,\xi})
\exp (i\Theta_{\mu,\nu,\xi}),
\end{equation}
where 
\begin{eqnarray}\label{Sdef1}
{\bf S}_{\mu,\nu,\xi}&=&
{\bf r}-{\bf v}_{\mu,\nu,\xi} t 
\end{eqnarray}
and
\begin{eqnarray}
\Theta_{\mu,\nu,\xi}
&=&\mbox{${\bf k}_{\mu,\nu,\xi}\cdot{\bf r}$}  -  \Omega_{\mu,\nu,\xi}t.
\end{eqnarray}
Here, ${\bf r}=\{m,n,p\}$ is the position vector, 
${\bf v}_{\mu,\nu,\xi}=\{v_x,v_y,v_z\}$ is the velocity, 
${\bf k}_{\mu,\nu,\xi}=\{\mu k_x,\nu k_y,\xi k_z\}$ is the
quasimomentum vector, and  $\Omega_{\mu,\nu,\xi}= \Omega_0(\mu
\omega_x+\nu \omega_y+\xi \omega_z)$ is the
frequency (chemical potential in BEC arrays).

By applying the quasicontinuum approximation
\cite{Neuper94,edward08,edward09-2D} we  obtain 
for   the first (F) harmonic $\mu=\nu=\xi=1$ of the expansion 
(\ref{ansatz1}) an equation which reads as
\begin{equation}\label{eqCartensian1}
\alpha_{x}\partial^2_{x} \chi_{F}+
\alpha_{y}\partial^2_{y}\chi_{F}+
\alpha_{z}\partial^2_{z}\chi_{F}
-\alpha_{0}\chi_{F}+
3 U \vert \chi_{F}\vert^2 \chi_{F}=0,
\end{equation}
where  $\chi_{F}=\chi_{1,1,1}$, and
\begin{eqnarray}\label{constants1}
\alpha_{0}&=&2[J_x\cos(k_x)+J_y\cos(k_y)+J_z\cos(k_z)](1-\Omega_0),\nonumber\\ 
\alpha_{x}&=&-J_x \cos(k_x),\nonumber\\ 
\alpha_{y}&=&-J_y \cos(k_y),\nonumber\\ 
\alpha_{z}&=&-J_z \cos(k_z).
\end{eqnarray}
In Eq. (\ref{eqCartensian1}) the coordinates $\{x,y,z\}={\bf
  S}_{1,1,1}$, where ${\bf S}_{\mu,\nu,\xi}$ is given in
Eq. (\ref{Sdef1}). In Eq. (\ref{constants1})   the constants
$\Omega_0$, $k_x$, $k_y$ and $k_z$ can be chosen as independent
parameters.

Since the coordinates $\{x,y,z\}$ are mutually independent,
Eq. (\ref{eqCartensian1}) can be integrated. So, finally, approximate
soliton solutions of Eq. (\ref{dnls1}) read as
\begin{eqnarray}\label{solsolution1}
&&\psi_{m,n,p}=e^{i\Theta_{1,1,1}}\sqrt{\frac{\alpha_{0}}{3 U}}\times\quad\nonumber\\
&&\prod\limits_{w=x,y,z}
[\delta_{\beta_w,1}2^{\frac{1}{6}}\sech^{\frac{1}{3}}(\frac{w}{L_{Bw}}) 
+\delta_{\beta_w,-1}\tanh^{\frac{1}{3}}(\frac{w}{L_{Dw}})],\nonumber\\
\end{eqnarray}
where $\delta_{\beta_w,\pm 1}$ is the Kronecker delta with
$\beta_w=\mbox{\rm sign}(\gamma_w\alpha_{0}/\alpha_{w})$ and $w=x,y,z$. 
In Eq. (\ref{solsolution1}) the soliton widths  read as 
\begin{eqnarray}
L_{Bw}&=&\sqrt{\gamma_w\alpha_{0}/\alpha_{w}} 
\end{eqnarray}
and
\begin{eqnarray}
L_{Dw}&=&\sqrt{-\gamma_w  \alpha_{0}/(2\alpha_{w})}, 
\end{eqnarray}
where the relation $\gamma_x+\gamma_y+\gamma_z=1$ should be satisfied.
Besides
\begin{eqnarray}
\Theta_{1,1,1}&=&k_x\, m + k_y\, n+ k_z\, p-\Omega_{1,1,1} t\\
\label{freq1}
\Omega_{1,1,1}&=&-2
\Omega_0[J_x\cos(k_x)+J_y\cos(k_y)+J_z\cos(k_z)],\quad\quad
\end{eqnarray}
and
\begin{eqnarray}
\label{veloc1}
{\bf v}_{1,1,1}&=&2 \{J_x \sin(k_x), J_y
\sin(k_y), J_z \sin(k_z)\}.
\end{eqnarray}

In Fig. \ref{fig1} it is shown an example of the motion of solitons
following from initial conditions given by Eq. (\ref{solsolution1}).

\begin{figure}
\centerline{\epsfxsize=7truecm \epsffile{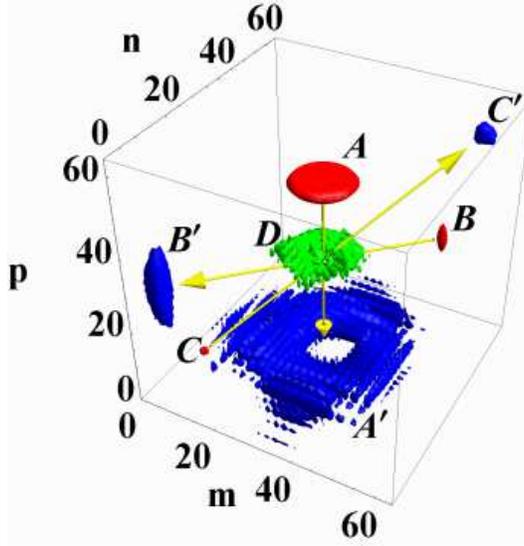}}
\caption{(Color online) Superposition of snapshots
similar as in Fig. \ref{fig1}, but with $\kappa=0.2$. Arrows in yellow show  the
path and direction of motion.
At $t=0$ [labels A, B, and C (red color)] the initial conditions
follow from  Eq. (\ref{solsolution2}), where (A) ${\bf k}=
\{\pi,\pi,-1.05\frac{\pi}{2}\}$, $\Omega_0=1.05$, $\nu=\mu=1$ [extended dark
vortex plotted in Fig. \ref{fig2}(b) (orange double-dot-dash line)
truncated with the function $0.5(1-\tanh(r-4)]$], (B)
  $\Omega_0=1.01$, $\nu=\mu=0$, and (C) $\Omega_0=1.05$,
$\nu=\mu=0$. At $t=10$ [label D (green 
  color)] the collision can be observed. And at $t=24$ [labels $\rm
  A^{\prime}$, $\rm B^{\prime}$, and $\rm C^{\prime}$ (blue color)]
the vortex and solitons after the collision can be observed. Other
parameters are given in Fig. \ref{fig1}.}
\label{fig3}
\end{figure}

We note that other soliton solutions with  embedded vorticity can be
calculated by using spherical coordinates $(r,\theta,\phi)$. In that
case Eq. (\ref{eqCartensian1}) can be written as
\begin{eqnarray}
&&\beta r^2 R(\vert R\vert^2-1)+\partial_r(r^2 R)+\nonumber\\
&&\partial_{\zeta}((1-\zeta^2)\partial_{\zeta}R)+
\partial^2_{\phi}R/(1-\zeta^2)=0,
\end{eqnarray}
where 
\begin{eqnarray}\label{Rfunc1}
R&=&\sqrt{3 U/\alpha_{0}}\chi_{F}, \\
\label{r2def1}
r^2&=&\alpha_{0}(x^2/\alpha_{x}+y^2/\alpha_{y}+z^2/\alpha_{z}),  \\
\zeta&=&\cos(\theta),\\
\theta&=&\textrm{tan}^{-1}[\sqrt{\alpha_{y}} x /(\sqrt{\alpha_{x}} y)],\\
\phi&=&\textrm{cos}^{-1}[\sqrt{\alpha_0}z /(\sqrt{\alpha_{z}}
  r)]. 
\end{eqnarray}
By using the ansatz $R=F(r)Y^{\mu}_{\nu}(\theta,\phi)$,
the equation for $R$ can be reduced to
\begin{equation}\label{spherical2}
\frac{d}{dr}\left(r^2\frac{d}{dr}f\right)+\beta r^2 f (f^2-1)-\nu(\nu+1)f=0.
\end{equation}
Here, $f=\vert Y^{\mu}_{\nu}\vert F(r)$ and $Y^{\mu}_{\nu}$ is a
spherical harmonic function of degree 
$\nu$ ($\nu \ge 0$) and order $\mu$ ($\vert \mu\vert \le \nu$). In
Eq. (\ref{spherical2}) the parameter $\beta=1$
($\beta=-1$) for  $\alpha_0\alpha_{w}>0$ ($\alpha_0\alpha_{w}<0$) with
$w=x,y,z$.

Finally we obtain the solution 
\begin{equation}\label{solsolution2}
\psi_{m,n,p}=\sqrt{\frac{\alpha_{0}}{3 U}}
\frac{Y^{\mu}_{\nu}(\theta,\phi)}{\vert
  Y^{\mu}_{\nu}(\theta,\phi)\vert}f(r)
\exp (i\Theta_{1,1,1}),
\end{equation}
where the  embedded vorticity is given by the order parameter $\mu$.
In Eq. (\ref{solsolution2}) the behavior of the function $f$
is governed by Eq. (\ref{spherical2}). Notice that for $\nu=0$ a solution of
Eq. (\ref{spherical2}) is $f=1$. For 
given $\nu$ and $\beta$ values in Eq. (\ref{spherical2}), other
unbounded and bounded solutions can be numerically calculated 
with a shooting method. Some of these $f$ solutions are plotted in
Fig. \ref{fig2}(b). Moreover, an  example of the motion of $\psi_{m,n,p}$
functions given in Eq. (\ref{solsolution2}) is shown in Fig. \ref{fig3}.

\section{Results}
The split-step Fourier method has been used 
to solve numerically Eq. (\ref{dnls1}) for a lattice size 
$\max(m\times n\times p)=64\times 64\times 64$
with periodic boundary conditions.

In Figs. \ref{fig1} and  \ref{fig3}
superpositions of snapshots at different time values of a 
soliton-soliton-soliton and  a vortex-soliton-soliton collision are
presented. The plots consist in  $\rho_{m,n,p}$  
isosurfaces (surface for which $\rho_{m,n,p}$ is constant). The initial
conditions in Figs. \ref{fig1} and \ref{fig3} [labels A, B, and C
  (color red)] follow from Eqs. (\ref{solsolution1}) and
(\ref{solsolution2}), respectively. 
The direction of motion of solitons and vortices is given by the 
the normalized velocity vector,
\begin{equation}
{\bf d}={\bf v}_{1,1,1}/\vert{\bf v}_{1,1,1}\vert,
\end{equation}
where ${\bf v}_{1,1,1}$ is given by Eq. (\ref{veloc1}).

In Figs. \ref{fig1} and  \ref{fig3} we observe 
that solitons and vortices when propagating undergo a 
self-defocusing instability. This inestability follows from the 
interplay between nonlinearity and discreteness.
Notice that nonlinearity tends to localize the wave
while the discrete diffraction tends to spread it out. For moving solitons 
the self-defocusing instability corresponds to the case when
diffraction effect is stronger than the nonlinear localization effect.
This self-defocusing process manifest itself as a 
soliton-shape broadening accompanied by exponential-like amplitude
decay. An example of  this amplitude decay can be observed in
Fig. \ref{fig2}(a), where the evolution of the  absolute  $\rho_{m,n,p}$
maximum ($\rho_{max}$) of the collision in Fig. \ref{fig1} is
plotted.

Notice that in Fig. \ref{fig2}(a) $\rho_{max}$
peaks when the collision  in Fig. \ref{fig1} [label D (green
  color)] occurs. After collision the solitons
further broaden in shape, as can be observed in  Fig. \ref{fig1}
[labels $\rm A^{\prime}$, $\rm B^{\prime}$, and $\rm
  C^{\prime}$ (blue color)]. Besides, for comparison in
Fig. \ref{fig2}(a)  the amplitude $\rho_{max}$ of the individual
solitons in the absence of collision is plotted. We can observe that
before and after collision 
the $\rho_{max}$ value coincides with that of individual
solitons. Moreover, we can observe that the peak amplitude during
collision is much higher than three times the amplitude of the
individual solitons. It is because during collision the nonlinear
effect becomes strong, so  a transitory self-focusing process takes
place.

The self-defocusing process in  solitons depends on their
form  and direction of motion, which in turn are governed by ${\bf k}$.
As in two dimensional systems \cite{quadartic2007,edward09-2D}, solitons 
moving along the diagonal directions, 
\begin{eqnarray}
{\bf d}=\left\{\begin{array}{l} 
\{\pm 1,\pm 1,\pm 1\}/\sqrt{3}\\  
\{\pm 1,\pm 1,0\}/\sqrt{2}\\ 
\{\pm 1,0,\pm 1\}/\sqrt{2} \\
\{0,\pm 1,\pm 1\}/\sqrt{2}
\end{array}
\right.\quad,
\end{eqnarray}
 are less prone to self-defocusing
than those moving along the main-axes directions,
\begin{eqnarray}
{\bf  d}=\left\{
\begin{array}{l}
\{\pm 1,0,0\}\\
\{0,\pm 1,0\}\\
\{0,0,\pm 1\} 
\end{array}
\right.\quad .
\end{eqnarray}
Notice that it is possible to control the
defocusing process so that solitons moving in different directions and
velocities can have the same decay rate. For example, solitons in
Fig. \ref{fig1} have similar decay rates, as shown in
Fig. \ref{fig2}(a). This can be easily achieved by first choosing  
with the help of ${\bf k}$ the wanted directions  and velocities of motion
and then with the help of $\Omega_0$, included in $\alpha_0$ in
Eqs. (\ref{constants1}), the same initial amplitudes. Here we use the
fact that the amplitude of a soliton is proportional to its decay
rate \cite{edward08,edward09-2D}. Notice that for similar decay
rates we obtain solitons  with different sizes and velocities, as shown
Figs. \ref{fig1} and \ref{fig3}.

In Fig.  \ref{fig3} we plot the initial soliton and vortex solutions
[labels A, B, and C (color red)] following from
Eq. (\ref{solsolution2}). In particular, the motion of a vortex 
shell
($\textrm{A}\rightarrow\textrm{D}\rightarrow\textrm{A}^{\prime}$
in Fig. \ref{fig3}) along its azimuthal axis (p axis) 
with vorticity $\mu=1$ is shown. The soliton dynamics is similar as in
Fig. \ref{fig1}, however small perturbations of the soliton shape are
observed after the collision (labels $\rm
  B^{\prime}$, and $\rm C^{\prime}$ in   Fig. \ref{fig3}).
The broadening and breaking-apart of the vortex in Fig. \ref{fig3}
is due to the modulation instability and can be also observed in the
absence of collisions. This  instability effect on vortices is more
pronounced for higher values of the vorticity $(\mu\ge 2)$.

We remark that the center of mass of both solitons
and vortices move with a constant velocity whose magnitude measured in
simulations is very well predicted by Eq. (\ref{veloc1}).
On the other hand, though not seen in Figs. \ref{fig1} and
\ref{fig3}, the presence of low-intensity radiation tails can be
detected for both soliton and moving vortices.

A question that emerges from the results above is how to mitigate
the self-defocusing effect. In order to tackle this problem we
investigate the effect of a magnification in the form of a 3D
thin-lens phase, i.e. 
\begin{equation}\label{lens1}
\psi_{m,n,p}\rightarrow \psi_{m,n,p}\exp(-i\, r^2/(4 T)). 
\end{equation}
In the present analysis $\psi_{m,n,p}$ 
solutions follow from Eq. (\ref{solsolution2}), where the radial
coordinate $r$ is given by Eq. (\ref{r2def1}). In Eq. (\ref{lens1}),
due to the discreteness of the  system, the parameter $T$ does not
correspond exactly to the  focal length, as it happens in continuous
systems.

\begin{figure}
\centerline{(a)\epsfxsize=7.0truecm \epsffile{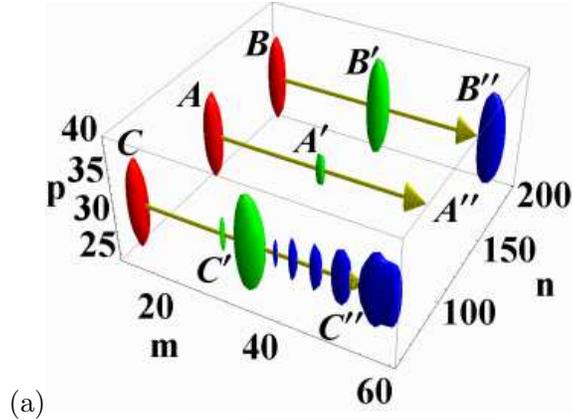}}
\centerline{(b)\epsfxsize=5truecm \epsffile{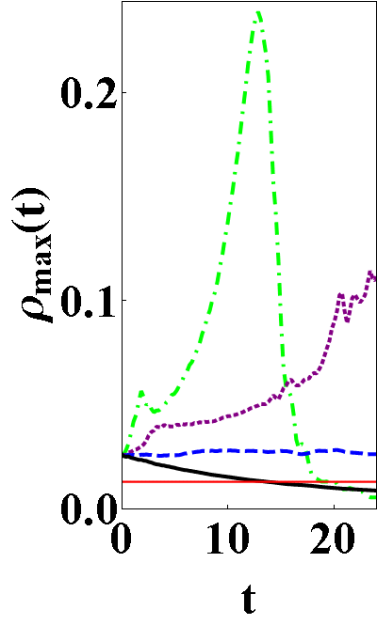}}
\caption{(Color online) a:  Superposition of snapshots of
 $\rho_{m,n,p}$ isosurfaces at different time values of
3 solitons with different thin-lens-phase factors. Here, the
isosurfaces are defined as the fix value $\kappa \operatorname*{max}_{m,n,p}\,
\rho_{m,n,p}(t=0)$ and $\kappa=0.5$. Arrows (in yellow) show  the
path and direction of motion. At $t=0$
[labels A, B, and C (red color)] the initial conditions
follow from  Eq. (\ref{solsolution1}), where (A) $T=\infty$, (B)
$T=0.2$, (C) $T=0.1$. At $t=12$ [labels $\rm
  A^{\prime}$, $\rm B^{\prime}$, and $\rm C^{\prime}$ (green color)]
and $t=24$ [labels $\rm A^{\prime\prime}$, $\rm B^{\prime\prime}$, and $\rm
  C^{\prime\prime}$ (blue color)] solitons move with different
maximums. Other parameters: ${\bf k}=0.95
\frac{\pi}{2}\{1,0,0\}$, $\Omega_0=1.001$, and $\nu=\mu=0$. 
b: Evolution of $\rho_{m,n,p}$ maximum of 
individual solitons. $T=\infty$ (Black solid line),  $T=0.2$ (blue
dashed line), $T=0.1$  (purple dotted line), and $T=0.05$ (green dot-dashed
line). The straight red line corresponds to the isosurface
level of the plots in  panel (a).}
\label{fig4}
\end{figure}

In Fig. \ref{fig4}(a) we consider three solitons moving in the
positive direction of the $m$ axis, and  their individual
 $\rho_{max}$ values are plotted in Fig. \ref{fig4}(b). In
Fig. \ref{fig4}(a) the transversal 
separation distance between the solitons is large enough to avoid
soliton-soliton interactions. The initial soliton forms, $\psi$
[Labels C, A,
B (red color) in Fig. \ref{fig4}(a)],  are identical to each
other but distinct from the $T$ value in the imposed thin-lens
phase. In order to compare the soliton behavior, 
the level value of the isosurfaces in
Fig. \ref{fig4}(a)  
was defined to be fixed to some initial value, as shown in
\ref{fig4}(b) (red line). So, solitons with $\rho_{max}$ value above
this level appear plotted in Fig. \ref{fig4}(a).
Notice, e.g., that in the soliton evolution
${\textrm A}\rightarrow {\textrm A}^{\prime}\rightarrow {\textrm
  A}^{\prime\prime}$ in 
Fig. \ref{fig4}(a) the soliton shape vanishes at $A^{\prime\prime}$
because its $\rho_{max}$ value [black line in Fig.  \ref{fig4}(b)] decays below the
level value [red line in Fig.  \ref{fig4}(b)] of the isosurface  for $t>12$.

Three different $T$ values have been considered in Fig. \ref{fig4}(a),
namely $T=\infty$, 0.2, and 0.1. The value $T=\infty$ (${\textrm A}\rightarrow{\textrm
  A}^{\prime}\rightarrow{\textrm A}^{\prime\prime}$)  corresponds to
the case where magnification is negligible. The value $T=0.2$
(${\textrm B}\rightarrow{\textrm B}^{\prime}\rightarrow{\textrm
  B}^{\prime\prime}$) corresponds to a case where 
for scale of time considered the soliton amplitude
remains nearly constant [see Fig. \ref{fig4}(b), blue dashed
line]. However, it is to remark that for  larger
time scales a self-defocusing process is observed and unavoidable.
The value  $T=0.1$ (${\textrm C}\rightarrow{\textrm
  C}^{\prime}\rightarrow{\textrm C}^{\prime\prime}$) corresponds to a
case where a focusing due to the thin-lens phase is immediately
followed by a self-focusing effect, as can be observed in 
Fig. \ref{fig4}(b) (purple dotted line).  Besides, a
strong radiation tail can be also observed in Fig. \ref{fig4}(a) ($\rm
C^{\prime\prime}$).

For further  comparison in Fig.  \ref{fig4}(b) the case of 
$T=0.05$ (green dot-dashed
line) is also plotted. This case is similar to the case $T=0.1$ where
the  first peak corresponds to a focusing due to the imposed
thin-lens phase and the second peak is due to a self-focusing
effect. Notice that after the second peak the soliton amplitude
strongly decays and a long radiation
tail, more in the fashion of Fig. \ref{fig4}(a) ($\rm
C^{\prime\prime}$), can be also observed.

\section{Conclusions}
We have studied, for the first time, the motion of
solitons and vortices in the 3D-DNLSE. In the tight binding limit this
is the most simple model for studying the effect of
nonlinearities in 3D crystal-like structures, such as 3D photonic crystals,
metamaterials and 3D BEC arrays. The analytical results, which
are supported by simulations, suggest that these moving excitations can
appear or be excited for any finite value of the model parameters. This
implies that solitary waves may play an important role in the design
of 3D crystal-like structures, such as cloaking devices. 
On the other hand,
we have observed that solitons survive collisions and their
modulation instability can be delayed. This supports the idea
that photonic crystals may be used as a 3D compact routing 
\cite{siliconCrystal09}
for optical pulses. From the practical standpoint, it remains
challenging the exact generation of those solitons studied
here. However, it does not preclude the fact that  moving excitations
observed in those systems can be analyzed as a superposition of
solitons and/or vortices.

\end{document}